\documentclass{mem}
\usepackage{natbib}\usepackage{txfonts}\usepackage{balance}
\usepackage{graphicx}
\usepackage[a4paper,breaklinks,dvipdfm]{hyperref}
\idline{75}{282}
\begin{document}
\def\teff{$T\rm_{eff }$}
\def\kms{$\mathrm {km s}^{-1}$}

\title{
Panchromatic fits to the Globular Cluster NGC 6366
}

   \subtitle{}

\author{
Fab\'iola \,Campos\inst{1},
S. O. \, Kepler\inst{1} 
\and C. \, Bonatto\inst{1}
          }

  \offprints{Fab\'iola Campos}

\institute{
$^1$ Instituto de F\'isica --
Universidade Federal do Rio Grande do Sul, Avenida Bento Gon\c{c}alves 9500,
91509-900 Porto Alegre, Rio Grande do Sul, Brasil\\
\email{fabiola.campos@ufrgs.br}
}

\authorrunning{Fab\'iola Campos }

\titlerunning{Panchromatic fits to the globular cluster NGC 6366}

\abstract{
We present panchromatic isochrone fits to the color magnitude data of the globular 
cluster NGC 6366, 
based on HST ACS/WFC and SOAR photometric data. Before performing the isochrone fits, 
we corrected 
the photometric data for differential reddening and calculated the mean ridge line
of the color magnitude diagrams. 
We compared the isochrones of Dartmouth Stellar Evolution Database and PAdova and 
TRieste Stellar Evolution Code (with microscopic diffusion starting on the main 
sequence). Based on previous determinations of the metallicity of this cluster 
we test it from [Fe/H]=-1.00 
to [Fe/H]= -0.50, and the age from 9 to 13 Gyrs. The uncertainties do not decrease 
when we fit simultaneous colors. We also find that the Dartmouth Stellar Evolution 
Database isochrones have a better fit in the sub giant branch and low main sequence 
than the PAdova and TRieste Stellar Evolution Code. Considering the most recent 
spectroscopic determination of the metallicity ([Fe/H]= -0.67), we find 
E(B-V)=0.67$\pm$0.02, (m-M)$_V$=14.94$\pm$0.05 and 11$\pm$2 Gyr for NGC 6366.
}
\maketitle{}

\section{Introduction}

Galactic globular clusters are considered the ideal laboratories for the 
study of stellar evolution, mainly because their color magnitude diagram (CMDs) 
have very specific characteristics. The stars, in most globular clusters, follow a
single isochrone, suggesting that they formed rougly at the same time and with the same
metallicity.\\
To obtain the astrophysical parameters of the globular cluster is necessary to fit models 
to the stars in the CMD. But this fitting is 
not simple, because there are two physical parameters that can vary to generate the models to be 
compared with the data, age and metallicity, and two other fitting parameters, extinction and distance. 
One method commonly used is the fit "by eye", but
it is necessary 
to be aware of the uncertainties 
in the construction of isochrones, since they increase the uncertainties in the fit.\\
Among the problems of the evolutionary models are the lack of a precise description of convection. 
Red giant stars have deep convective envelope causing a large uncertainty in the structure. 
Another problem is that stars 
lose mass (higher for the most massive stars) in the form of a stellar wind and this 
loss increases several 
orders of magnitude for stars that already left the main sequence. Predicting the rate of mass loss 
theoretically is very difficult, and what the evolutionary models use are the values of 
mass loss consistent with observations of stars that are at a similar stage.
This value depends heavily on metallicity, and this dependence is difficult of 
measurement, causing uncertainties. 
It is still necessary to account the uncertainties associated with opacity tables, especially
when dealing with molecular opacities. This effect is important not only in the cooler giant 
stars, but also in the lower main sequence.\\
According to \citet{bolte95},
if the brightness of the main sequence turn off is used to determine the age of 
globular clusters, the only source of uncertainty significantly large is the determination of
distance and an uncertainty of 25\% in distance generates an uncertainty of
22\% in age.\\
Attempting to decrease the uncertainties of isochrone fittings to globular clusters,
we performed panchromatic isochrone fits of the globular cluster NGC 6366, 
based on archival HST ACS/WFC and our own 4.1m SOAR photometric data. We compared the 
isochrones of Dartmouth Stellar
Evolution Database [DSED - \citet{dotter08}] and PAdova and TRieste Stellar Evolution Code 
[PARSEC - \citet{bressan12}], models with microscopic diffusion starting on the main sequence
\citep{jofre}.

\section{Observations}

The optical data of NGC 6366 discussed in this work was observed with the SOAR telescope. 
The images were centered on the cluster core and the exposure 
times for each filter were divided as follows: 6x(1800s) for U; 5x(30s), 
2x(300s) and 1x(1800s) for B; and 5x(30s), 2x(300s) and 2x(1800s) for V . The photometry  
was obtained with psf fitting using DAOPHOT \citep{stetson}. To obtain the magnitudes in the standard photometric 
system, stars belonging to the cluster itself were used. the standard magnitudes for these stars were obtained through 
the catalog of standard stars of Peter 
Stetson (\url{http://www3.cadc-ccda.hia-iha.nrc-cnrc.gc.ca/community/STETSON/standards/}).
The photometric data of HST ACS/WFC was obtained from \url{http://www.astro.ufl.edu/~ata/public_hstgc/}.
This data was obtained as part of the HST treasury program \textquotedblleft{An ACS Survey of 
the Galactic Globular Clusters}\textquotedblright [GO10775 P.I. Ata Sarajedini, \citep{sarajedini}].
The images were centered at the cluster core
and consists of 10x(140s) and 1x(10s) exposures in the F606W and F814W filters.

\section{Data Analysis}

\begin{figure*}[t!]
\resizebox{\hsize}{!}{\includegraphics[clip=true]{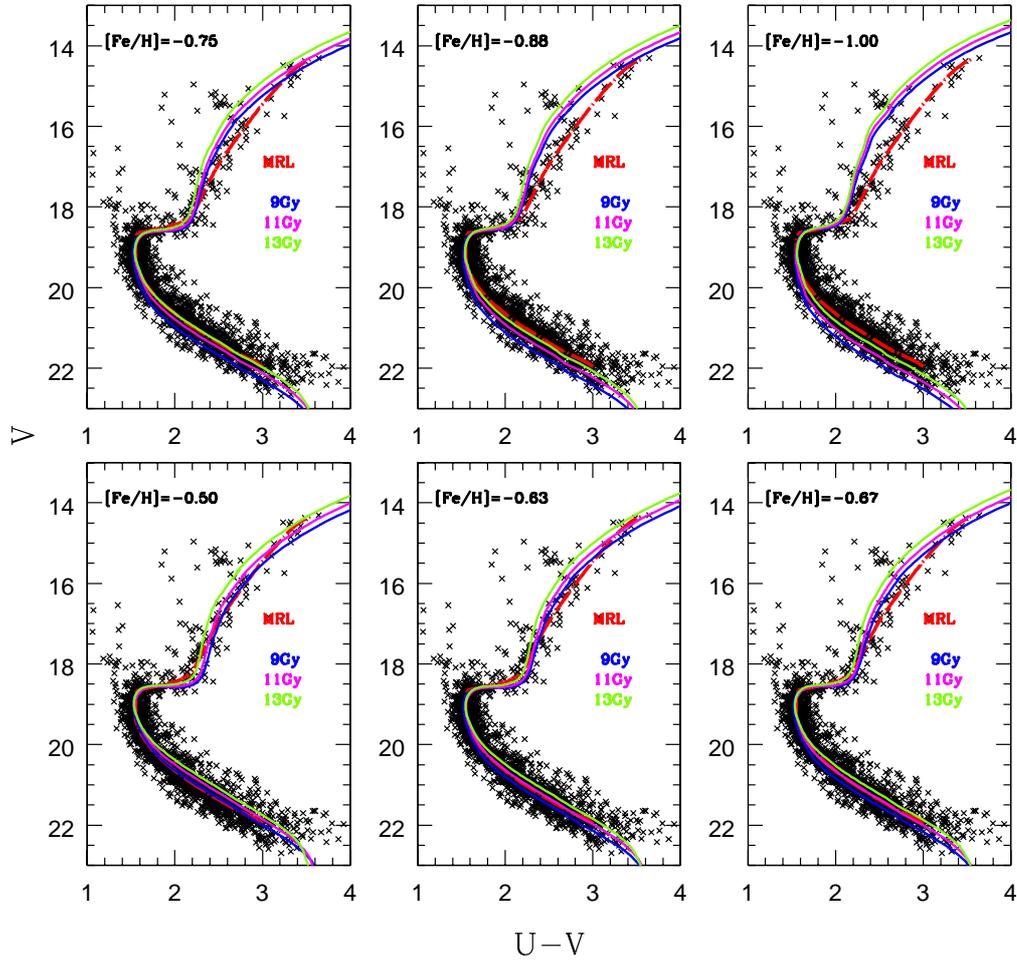}}
\caption{\footnotesize
Fitting of the DSED to the mean ridge line (red dash-dotted line) of NGC 6366, at U-V color, 
considering 
all the range of metallicity determinations for this cluster and ages of 9 Gyr (blue line), 
11 Gyr (magenta line) and 13 Gyr (green line).
The best fit is found at metallicity [Fe/H]=-0.67, value consistent with the
most recent spectroscopic determination by \citet{dacosta95}.
}
\label{all}
\end{figure*}

To perform the analysis, the first step involves 
correcting the photometric data for 
differential reddening by dividing the field of view across the cluster 
in a regular cell grid, then extract the Hess diagram 
from each cell, shifting it, along the reddening vector, until it matches
the mean diagram [Bonatto, Campos \& Kepler (2013), MNRAS, submitted].\\
To take into account the effects of binarity and also the scattering 
photometry, before performing any fitting to the CMDs, we calculated 
the mean ridge line of each CMD.
The fitting of the DSED and PARSEC models to the mean ridge line of the three colors 
(VxB-V, VxU-V and F606WxF606W-F814W) was performed considering all the range 
of metallicity previously determined for this cluster (-1.0$<$[Fe/H]$<$-0.5, 
\citet{ferro} and references therein) and ages ranging from 9 Gyr to 13 Gyr. 
In Fig. \ref{all} we show the fitting of DSED models to the mean ridge line at U-V color. 
It is not difficult to notice, by looking at the low main sequence and the sub giant 
branch that, as the 
metallicity decreases, the models have a better fit to the data, until [Fe/H]=-0.67, 
when the best fit is found, this value is consistent with the spectroscopic determination
by \citet{dacosta95}. As metallicity continues to decrease, the models no longer
fit the data.

\begin{figure}[]
\resizebox{\hsize}{!}{\includegraphics[clip=true]{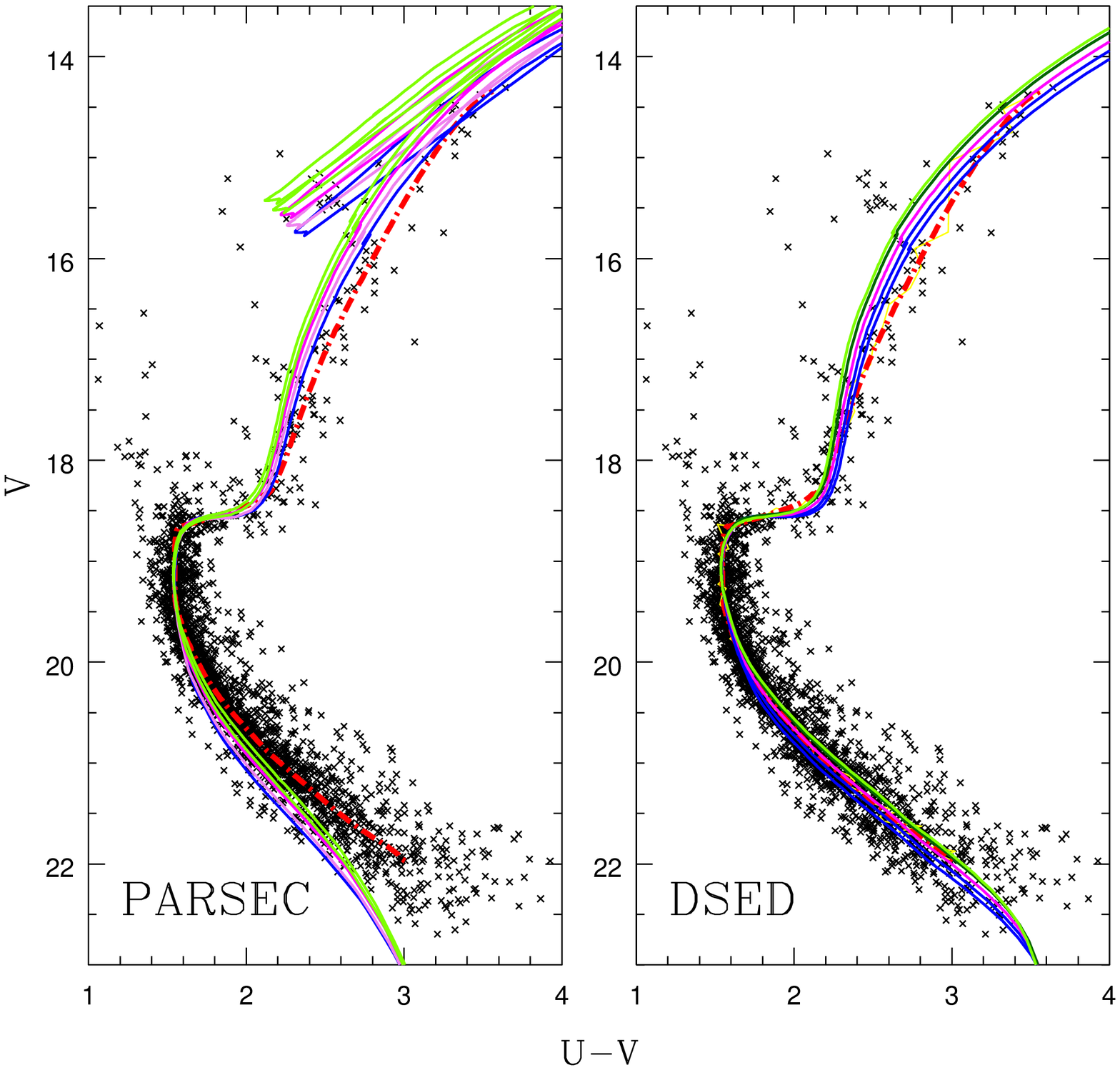}}
\resizebox{\hsize}{!}{\includegraphics[clip=true]{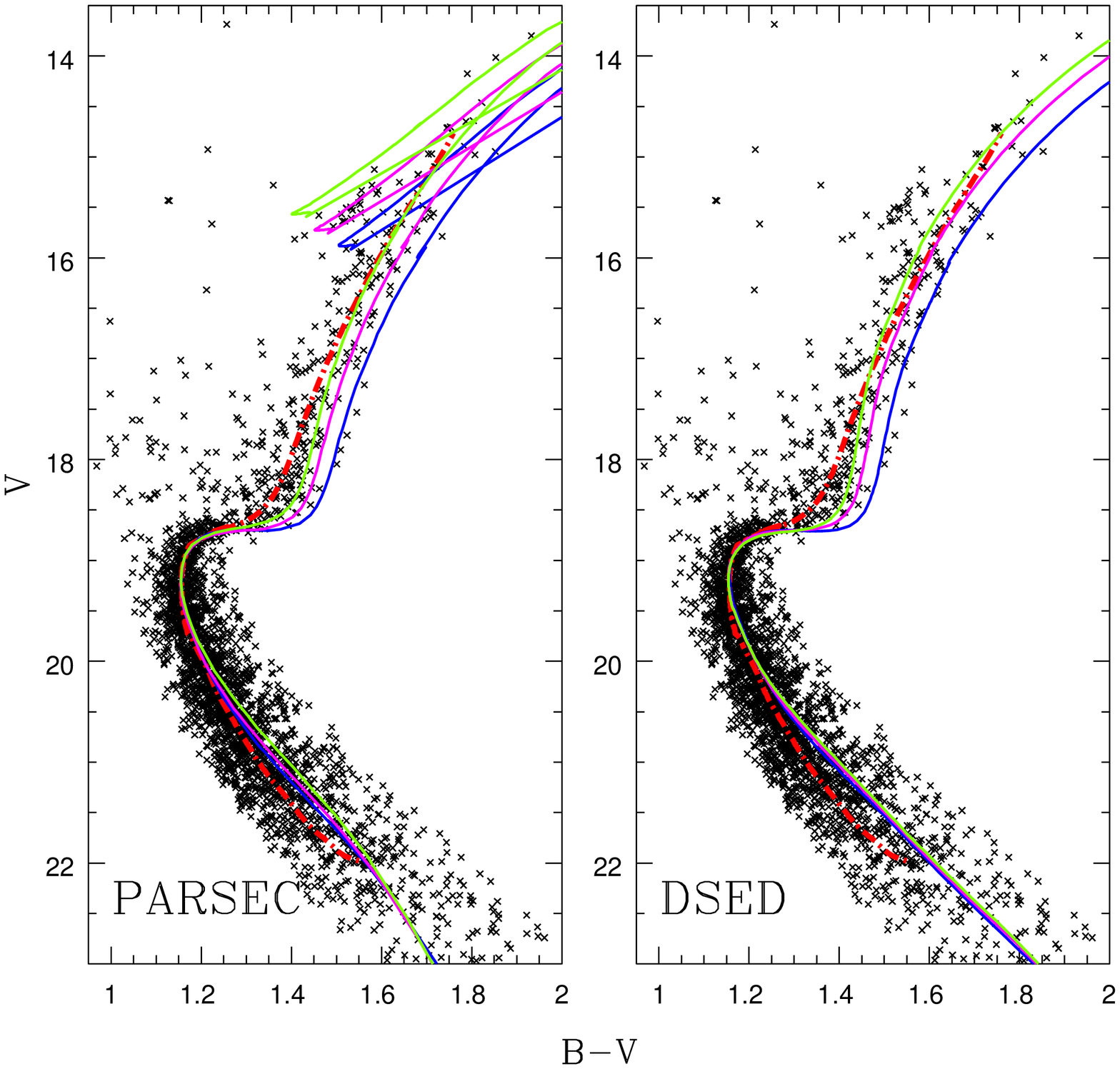}}
\resizebox{\hsize}{!}{\includegraphics[clip=true]{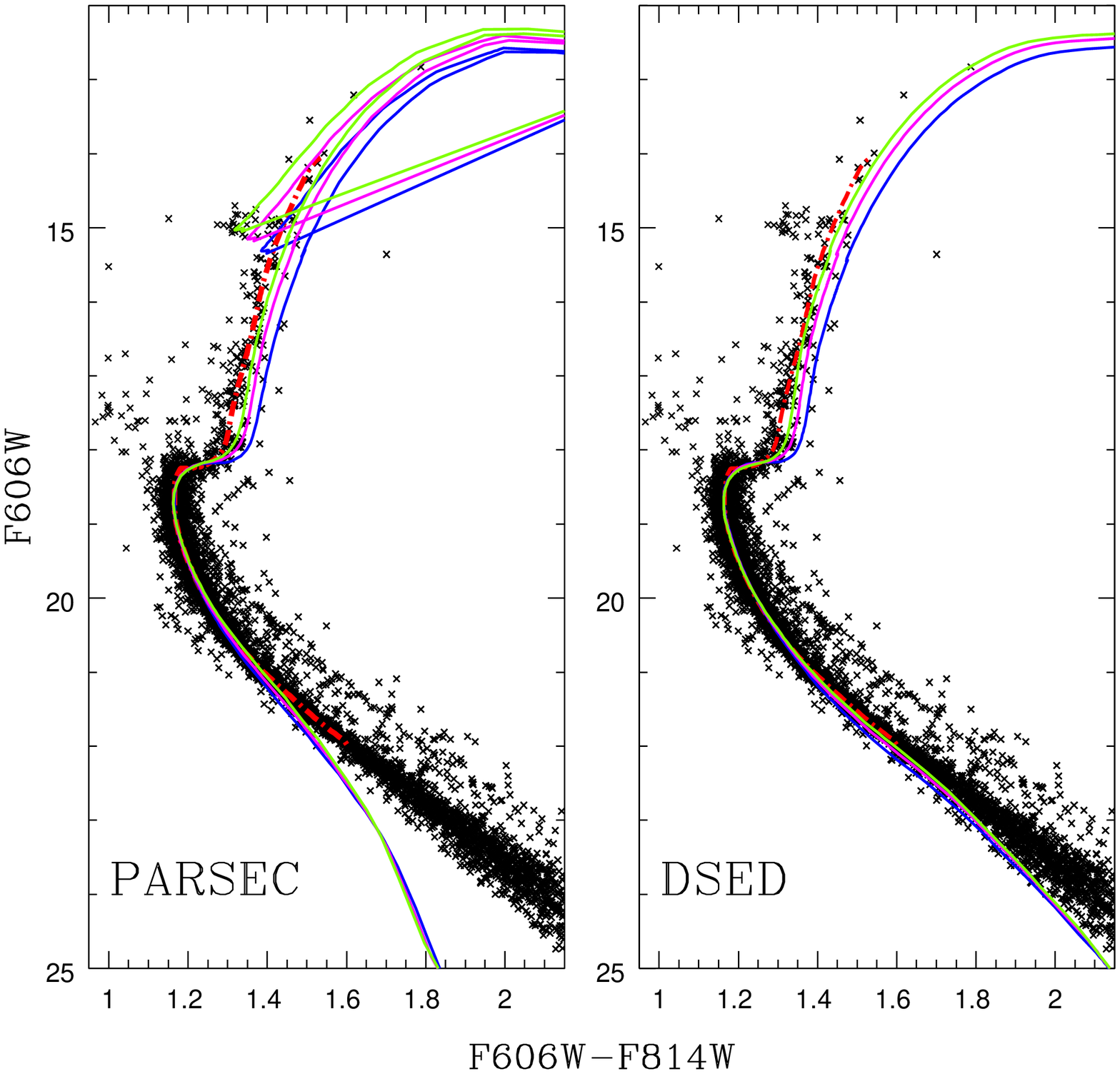}}
\caption{
\footnotesize
Fitting of PARSEC and DSED models to the mean ridge line (red dash-dotted line) 
of the SOAR and ACS/WFC data, with [Fe/H]=0.67$\pm$0.07 \citep{dacosta95} 
and ages of 9 Gyr (blue line), 11 Gyr (magenta line) and 13 Gyr (green line). 
The uncertainties do not decrease when we fit simultaneous colors, because a
single model does not fit the three colors simultaneously. The DSED
models have a better fit in the Hertzprung gap and the lower main sequence than the
PARSEC models.
}
\label{fits}
\end{figure}

The fitting for the three colors with [Fe/H]=-0.67$\pm$0.07 \citep{dacosta95} (Fig. \ref{fits}) 
shows that while the best fit to U-V color occurs at age 9 Gyr, for B-V color
the best fit is 
at 11 Gyr and for F606W-F814W color the best fit is found with 13 Gyr. In other words, a single 
model does not fit the three colors simultaneously, and the uncertainties do not decrease
when simultaneous colors are fitted. Indicating that the evoluionary models still have unsolved problems, 
such as photometric zero points and opacity tables.\\
Another interesting point to notice in Fig. \ref{fits} is the fact that DSED models have a better 
fit at the lower main sequence and at the sub giant branch than the PARSEC models; therefore
we use the DSED fittings to determine the parameters of the cluster. 
Before we find the relations between the colors, to obtain the parameters, we determined the 
total to selective extinction for stars belonging 
to this cluster directly, following \citet{duca03}, 
and we find R$_V$=3.06$\pm$0.14. We also used \citet{ccm} 
relations to estimate E(B-V)=1.02E(F606W-F814W) and E(B-V)=0.57E(U-V). 
We find the following parameters for NGC 6366:

E(B-V)= 0.69$\pm$0.02(int)$\pm$0.04(ext)

(m-M)$_V$= 15.02$\pm$0.07(int)$\pm$0.13(ext)

Age= 11$\pm$2 Gyr.

\section{Conclusions}

The uncertainties of isochrone fittings to NGC 6366 do not
decrease when we fit multiple colors, due to the fact that a single model
does not fit three colors simultaneously. It indicates that the isochrone models may
still have problems that remain unsolved, possibly photometric zero points and opacity tables.
Other important point is that the DSED models 
have a better fit to the data than PARSEC models, mainly in the sub giant branch and the 
low main sequence, the last one is possibly realted to the equation of state adopted by \citet{dotter08}
for stars with low mass.\\
With the total to selective relation R$_V$=3.06$\pm$0.14, determined for stars belonging 
to this cluster, we estimated E(B-V)=1.02E(F606W-F814W) and E(B-V)=0.57E(U-V). With that we 
determined the parameters of NGC 6366 as being: E(B-V)= 0.69$\pm$0.02(int)$\pm$0.04(ext); 
(m-M)$_V$= 15.02$\pm$0.07(int)$\pm$0.13(ext) and Age= 11$\pm$2 Gyr.

\begin{acknowledgements}
Finacial support for this research comes from  National 
Council for Scientific and Technological Development (CNPq) 
and PRONEX-FAPERGS/CNPq.
\end{acknowledgements}

\bibliographystyle{aa}

\end{document}